\documentclass[twocolumn]{aastex631}

\usepackage{amsmath}
\usepackage{makecell}
\usepackage{comment}
\usepackage{natbib}
\usepackage{multirow}
\usepackage{booktabs}
\usepackage{epsfig}
\usepackage{ulem}
\usepackage{rotating}
\usepackage{graphicx}
\usepackage{url}
\usepackage{xurl} 
\usepackage{xcolor}
\usepackage{amsmath}
\usepackage{color}
\usepackage{savesym}
\let\tablenum\relax
\usepackage{siunitx}
\usepackage[normalem]{ulem}
\usepackage{appendix}
\usepackage{enumerate}
\usepackage{gensymb}
\usepackage{multirow}
\usepackage{hyperref}
\usepackage{tikz}
\usepackage{dashrule}
\usepackage{array}    
\usepackage{float}

\usepackage{longtable}
\usepackage{afterpage}
\usepackage{threeparttable}
\usepackage{tabularx}

\DeclareSIUnit\parsec{pc}
\DeclareSIUnit\h{\textit{h}}
\let\tablenum\relax



\begin{document}
    \title{Measuring the Distances to Asteroids from One Observatory in One Night with Upcoming All-Sky Telescopes}

  \author[0000-0002-9561-9249]{Maryann Benny Fernandes\footnotemark[1]}
    \affiliation{Department of Electrical and Computer Engineering, Duke University, Durham, NC 27708, USA}

    \author[0000-0002-4934-5849]{Daniel Scolnic\footnotemark[2]}
    \affiliation{Department of Physics, Duke University, Durham, NC 27708, USA} 
    \affiliation{Department of Electrical and Computer Engineering, Duke University, Durham, NC 27708, USA}

    \author[0000-0001-8596-4746]{Erik Peterson}
    \affiliation{Department of Physics, Duke University, Durham, NC 27708, USA}

    \author[0000-0002-0291-4522]{Chengxing Zhai}
      \affiliation{Division of Emerging Interdisciplinary Areas, The Hong Kong University of Science and Technology, Clear Water Bay, Hong Kong}

    \author[0000-0003-2534-673X]{Tyler Linder}
      \affiliation{Planetary Science Institute, Tucson, AZ 85719}

      \author[0000-0002-5389-7961]{Maria Acevedo}
    \affiliation{Department of Physics, Duke University, Durham, NC 27708, USA}

    \author[0000-0002-5060-3673]{Daniel Reichart}
      \affiliation{Department of Physics and Astronomy, University of North Carolina at Chapel Hill, Chapel Hill, NC 27599, USA} 

\footnotetext[0]{Corresponding Authors:\\ 
a - \href{mailto:maryannbenny.fernandes@duke.edu}{maryannbenny.fernandes@duke.edu},\\
b - \href{mailto:daniel.scolnic@duke.edu}{daniel.scolnic@duke.edu}.}

\begin{abstract}
Upcoming telescopes like the Vera Rubin Observatory (VRO) and the Argus Array will image large fractions of the sky multiple times per night yielding numerous Near Earth Object (NEO) discoveries. When asteroids are measured with short observation time windows, the dominant uncertainty in orbit construction is due to distance uncertainty to the NEO. One approach to recover distances is from \textit{topocentric parallax}, which is a technique that leverages the rotation of the Earth, causing a small but detectable sinusoidal additive signal to the Right Ascension (RA) of the NEO following a period of 1 day. In this paper, we further develop and evaluate this technique to recover distances in as quickly as a single night. We first test the technique on synthetic data of 19 different asteroids ranging from $\sim0.05 \,\text{AU}$ to $\sim2.4 \,\text{AU}$. We modify previous algorithms and quantify the limitations of the method, recovering distances with uncertainties as low as the $\sim1.3\%$ level for more nearby objects ($\lesssim$ 0.3 AU) assuming typical astrometric uncertainties. We then acquire our own observations of two asteroids within a single night with $\sim0.1''$ uncertainties on RA, and we find we are able to recover distances to the $3\%$ level. We forecast likely scenarios with the VRO and the Argus Array with varying levels of astrometric precision and expected pointings per night. Our analysis indicates that distances to NEOs on the scale of $\sim0.5$ AU can be constrained to below the percent level within a single night, depending on spacing of observations from one observatory. In a follow-up paper, we will compare these constraints with synchronous and asynchronous observations from two separate observatories to measure parallax even more efficiently, an exciting and likely possibility over the upcoming decade.
\end{abstract} 


\section{Introduction}
\label{sec:section1}
The detection and precise orbit determination of Near-Earth Objects (NEOs), objects with orbits that can come within 1.3 AU of the Sun,\footnote{\url{https://ssd.jpl.nasa.gov/sb/neos.html.}} are crucial for planetary defense and space exploration initiatives. NASA's Planetary Defense Coordination Office, and their NEO observation program\footnote{\url{https://science.nasa.gov/planetary-defense-neoo/}.}, funds surveys such as the Catalina Sky Survey (CSS) \citep{CSS}, Panoramic Survey Telescope and Rapid Response System (Pan-STARRS) \citep{Pan-STARRS}, and the Asteroid Terrestrial-impact Last Alert System (ATLAS) \citep{ATLAS_asteroids}, leading to the discovery of thousands of new NEOs (asteroids and comets) annually \citep{NEO's}. 

The follow-up observation time window for NEOs may be very short depending on the size of or distance to the asteroid. Historically, orbital determination of NEO utilized classical Gaussian and Laplacian methods \citep[e.g.,][]{Teets1999,Branham2005}, where the dominant uncertainties are from distance constraints to the NEO. One method that enables measurements of distances to NEOs quickly, is the \textit{topocentric parallax} method which has been extensively studied in \citet{Zhai2022} (hereafter Z22). This method leverages the apparent motion of NEOs as seen from either different observatory locations or the same observatory at various times due to Earth's rotation, allowing for more precise distance measurements more rapidly. The technique is effective for NEOs of various sizes, that are observed over short time windows, or short `observation arcs' -- ranging between a few hours to a couple days.

Z22 parameterize the uncertainty in orbital constraint in terms of the product of the observation arc length in days (\textit{T}) with distance (\(\Delta \)). When the product is small, \( T \Delta \lesssim 1\) day AU (when either the observation arc length or distance is small, or both), a precise and accurate determination of the parallactic distance can significantly improve the orbital constraint. Additionally, Z22 show that to effectively leverage the parallax technique for a single observatory, the hour angle (the difference between the local sidereal time and RA of the observations) must vary. 

There are other new parallax approaches that have been proposed to obtain quick and precise distance measurements using short observational arcs. In particular, \citet{Heinze2015} calculated the precise distances to 197 main belt asteroids (MBA's) at the $1-3\%$ level by using the Rotational reflex velocity (RRV) method. These MBA's were observed for only two nights from a single location. The RRV technique determined the distances to the asteroids by utilizing the Earth's rotation and the angular reflex motion of the asteroids, a velocity that is initiated into an asteroids motion due to the Earth's axial rotation. \citet{Guo2023} refined the \citet{Heinze2015} RRV method for use on Near Earth Asteroids (NEAs), such as, the observations of the Potentially Hazardous Asteroid (PHA) (99942) Apophis over two successive nights and its synthetic ephemerides data (set of positions and motions) from the NASA JPL Horizons System (hereafter \textit{Horizons}).\footnote{\url{https://ssd.jpl.nasa.gov/horizons/app.html\#/.}} The  resultant mean relative error between the observations and the synthetic data was  $\sim0.08\%$. \citet{Guo2023} also tested the RRV refinement on asteroids from each of the four NEA groups. The observations for these asteroids were selected such that the \textit{Horizons} ephemerides data was taken on the night of the asteroids discovery and its consecutive night, resulting in a $\textless$ $1\%$ relative error. \citet{Alvarez2012} used the diurnal parallax, a method similar to topocentric parallax to calculate the distances to the asteroids. The authors show that with this technique one can reach a distance accuracy at the 5\% level. While these techniques can reach high precision, in this paper, we focus on the topocentric parallax technique in Z22 as it may lend itself more easily to future all-sky surveys with multiple observations from a single night, rather than scheduled observations over multiple nights. We also note that recently, \cite{Kruk22} applied this technique to asteroid streaks found in archival \textit{Hubble Space Telescope} (\textit{HST}) images, and leveraged the fast orbit of \textit{HST} to determine the asteroid's parallax.

In Section~\ref{sec:section2}, we further develop the technique from Z22 due to the imminent arrival of wide-area telescopes with fast cadences. We apply this technique on a diverse set of asteroid ephemerides data from \textit{Horizons} and characterize the effectiveness of this approach. In Section~\ref{sec:section3}, we obtain our own data of two asteroids and recover distances. In Section~\ref{sec:section4}, we predict the constraint on distances from observations of future all-sky surveys like the Vera Rubin Observatory and the Argus Array. The predictions will offer valuable insight into how the NEO community can enhance orbit determination efforts on a larger scale, supporting both planetary defense strategies and NASA's space exploration goals.

\section{Methodology}
\label{sec:section2}

We review and further develop Z22's general method, analyze synthetic data to test on, and propose improvements and limitations to the approach. 

\subsection{Determining Distances}
To determine distances to asteroids within a single night, we must analyze RA measurements, which are sensitive to Earth's rotation, unlike Declination (DEC). By plotting RA over time, the asteroid's motion demonstrates a sinusoidal signal due to Earth's rotation in addition to the asteroid's linear motion. This relative motion can be approximated as a leading order approximation and expressed relative to the center of the Earth:
\begin{equation}
{RA_{\text{pred}}}  = \mathcal{A} \sin(2\pi t + \phi) + \alpha + \beta t.
\label{eq:linear}
\end{equation}

 \noindent Here, $RA_{\text{pred}}$ is the predicted RA value; \( t \) is the Julian Date; \( \mathcal{A}\) represents the amplitude of the sinusoidal motion observed from the Earth's rotation, which is also the parallax amplitude; \( \phi \) is a phase shift parameter; \( \alpha \) is a constant offset, and \( \beta \) represents the linear motion of the asteroid. Since the period (frequency) for the sine model is due to the Earth's rotation, we fixed the period to 1 day. By fitting the model to the array of RA data against time, we determine the parameters \( \mathcal{A} \), \( \phi \), \( \alpha \), and \( \beta \) by minimizing the residuals using a non-linear least squares fit with chi-squared ($\chi^2$) values expressed as: 

\begin{equation}
 \chi^2 = \sum \left (\frac{RA_{\text{meas}}-RA_{\text{pred}}}{\sigma_{\text{meas}}}\right)^2,
\label{eqn:chi^2}
\end{equation}

\vspace{0.5cm}

 \noindent where $RA_{\text{pred}}$ is from Eq.~\ref{eq:linear} consisting of all the optimization parameters, $RA_{\text{meas}}$ is the measured RA, and $\sigma_{\text{meas}}$ is the uncertainty in RA from the astrometric measurements of the asteroid. We assume the astrometric uncertainties are independent and that Eq.~\ref{eqn:chi^2} could be replaced with a covariance matrix when accounting for correlated measurement uncertainties like differential chromatic refraction (Lee \& Acevedo et al. \citeyear{Lee2023}). 

The amplitude from Eq.~\ref{eq:linear} is related to the angular displacement caused by the rotation of the Earth. Using the Earth's radius ($R_{\text{Earth}}$ = 6371 \,$\mathrm{km}$), we convert to a physical distance; serving as an intermediate step to obtain the distance to the asteroid from the center of the Earth:
\begin{equation}
   \text{Distance}_{\text{km}} = \frac{R_{\text{Earth}}}{\mathcal{A}_{\text{radians}}}.
   \label{eq:parallax}
\end{equation}

  \noindent We must also consider the latitude of the observatory because it affects the angle of parallax, which is influenced by the rotation of the earth and its curvature, producing an angular displacement between observations. Thus, Eq.~\ref{eq:parallax} is adjusted by multiplying the parallax distance with the cosine of the observatory's latitude as seen in Eq.~\ref{eqn:cos}, to accurately recover the distance to the asteroid:
\begin{equation}
   \text{Asteroid Distance} = \frac{\text{Distance}_{\text{km}} \times \cos(\text{latitude})}{\cos(\text{DEC})}.
   \label{eqn:cos}
\end{equation}

 \noindent The $\cos(\text{DEC})$ term is the cosine of the Declination angle of the object, and has the strongest effect on the parallax estimate when the object is near the celestial equator $(0^{\circ})$ and is negligible at the poles $(\pm 90^{\circ})$. For a Declination of $90^{\circ}$, this equation would be singular and cannot be used. Finally, we convert the asteroid's distance to astronomical units (AU), where \( 1 \, \text{AU} = 1.496 \times 10^8 \, \text{km} \). 

We allow for one further refinement to this technique. To account for a longer observation arc, we propose adding a quadratic term ($\varphi t^2$) to Eq.~\ref{eq:linear}, such that Eq.~\ref{eq:poly} addresses the non-linear motion of the asteroid:   
\begin{equation}
{RA_{\text{pred}}} = \mathcal{A} \sin(2\pi t + \phi) + \alpha + \beta t +\varphi t^2.
\label{eq:poly}
\end{equation}

\begin{figure*}
    \centering
            \includegraphics[width=2.1\columnwidth]{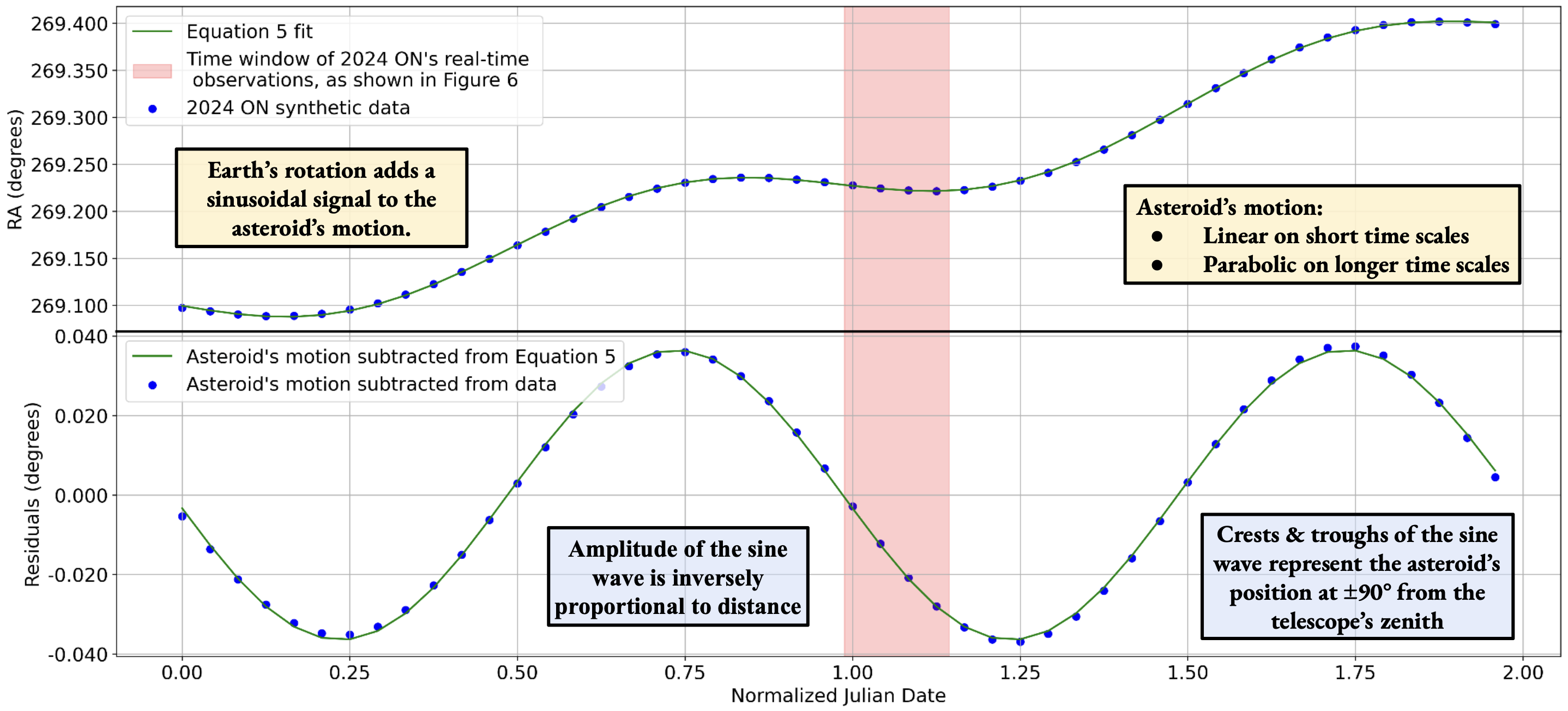}
    \caption{We obtain synthetic data for asteroid 2024 ON from \textit{Horizons}. The ephemerides are generated by  numerically integrating the orbits of celestial objects to their real-time observations \citep{Ryan2014}. (row 1) The raw RA positions as a function of time where the dots indicate the 48 observations taken from Sept 5 - 6, 2024 at every hour. The first-order behavior is due to the asteroid's motion. The smaller sinusoidal behaviour is due to the rotation of the Earth. (row 2) The remaining sinusoidal behaviour after subtracting out a second-order polynomial fit is shown in blue dots. The period of the sine curve due to the rotation of the Earth is 1 day and indicated by the green line. The amplitude of the sine curve is used in Eq.~\ref{eq:parallax} to measure the distance to the asteroid as seen in Eq.~\ref{eqn:cos}. The amplitude changes with time as the distance to the asteroid changes with time. The highlighted red region in both rows represent the real-time observations of 2024 ON, as discussed in Section~\ref{sec:section3} and as seen in Fig.~\ref{fig:coma}.}
    \noindent\hdashrule[0.5ex]{\linewidth}{0.1mm}{0.1mm}
    \label{fig:stepstep}
\end{figure*}

\begin{figure}[!htbp]
    \centering
    \includegraphics[width=\columnwidth,height=6.5cm]{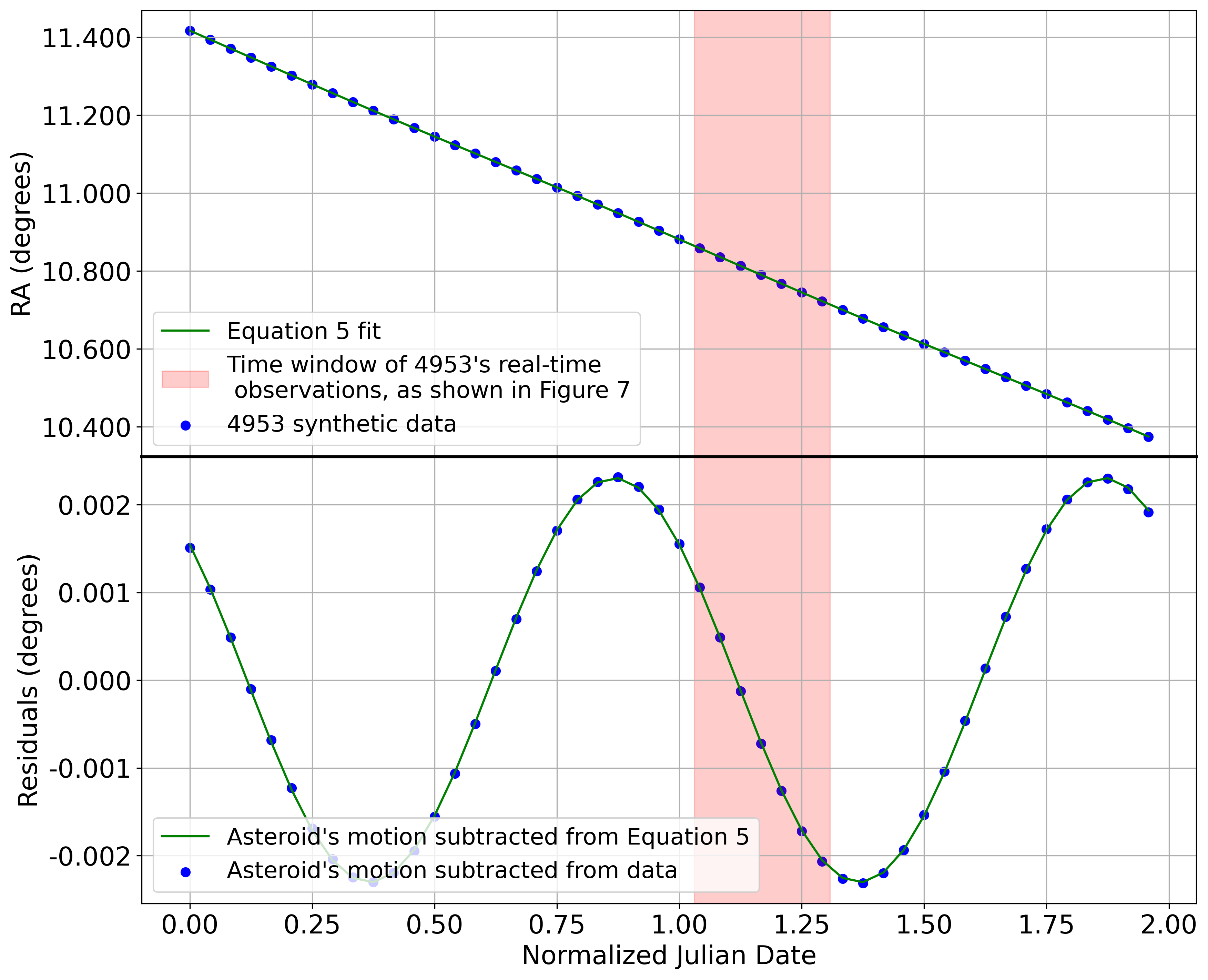}
    \caption{Similar to Fig.~\ref{fig:stepstep}. (row 1) We obtain synthetic data for 4953 from \textit{Horizons}. The blue dots indicate the RA observations from October 30 to 31, 2024 at every hour, totalling 48 observations. We fit Eq.~\ref{eq:poly} to the data as shown in the green line. (row 2) The residuals after subtracting out the asteroids motion from Eq.~\ref{eq:poly} is shown in the green line and from the data is shown in blue dots. The highlighted red region in both rows represent the real-time observation of 4953 as seen in Fig.~\ref{fig:4953_fit}.}
    \label{fig:stepstep4953}
\end{figure}

\begin{figure*}
    \centering
            \includegraphics[width=2.1\columnwidth]{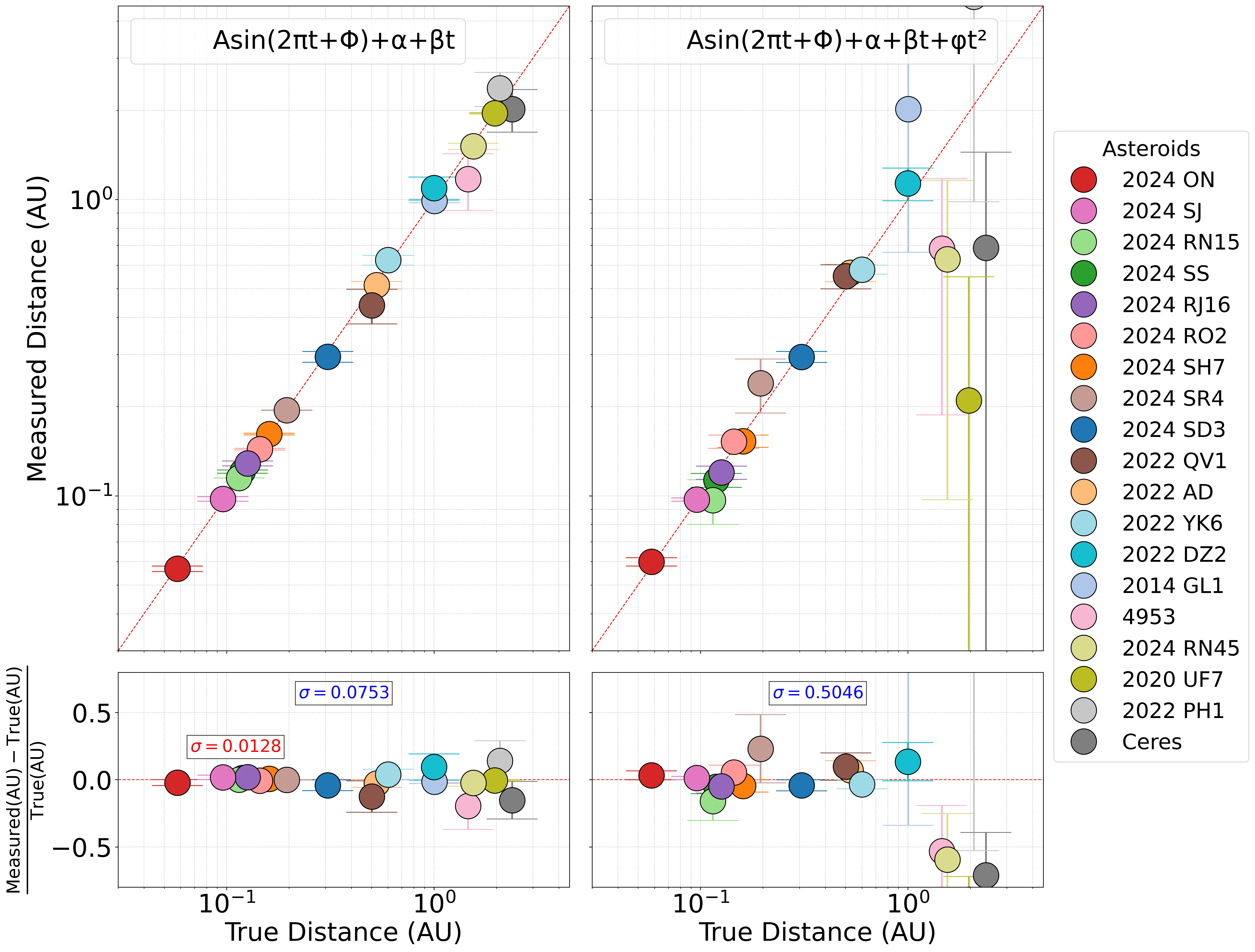}
    \caption{We recover distances to 19 asteroids, with each asteroid having 6 observations taken during one night and every observation separated by 45 minutes on September 5-6, 2024, from 23:43:00 to 03:28:10. (Top) Measured distance (AU) versus True distance (AU) for all 19 asteroids considered in this analysis. The left panel shows results from using Eq.~\ref{eq:linear} and the right panel from Eq.~\ref{eq:poly}. (Bottom) For the 19 asteroids the fractional residual is defined in Eq.~\ref{eq:fractional-distance-scatter}.}
    \noindent\hdashrule[0.5ex]{\linewidth}{0.1mm}{0.1mm}
    \label{fig:One_day}
\end{figure*}

\begin{figure*}
    \centering
            \includegraphics[width=2.1\columnwidth]{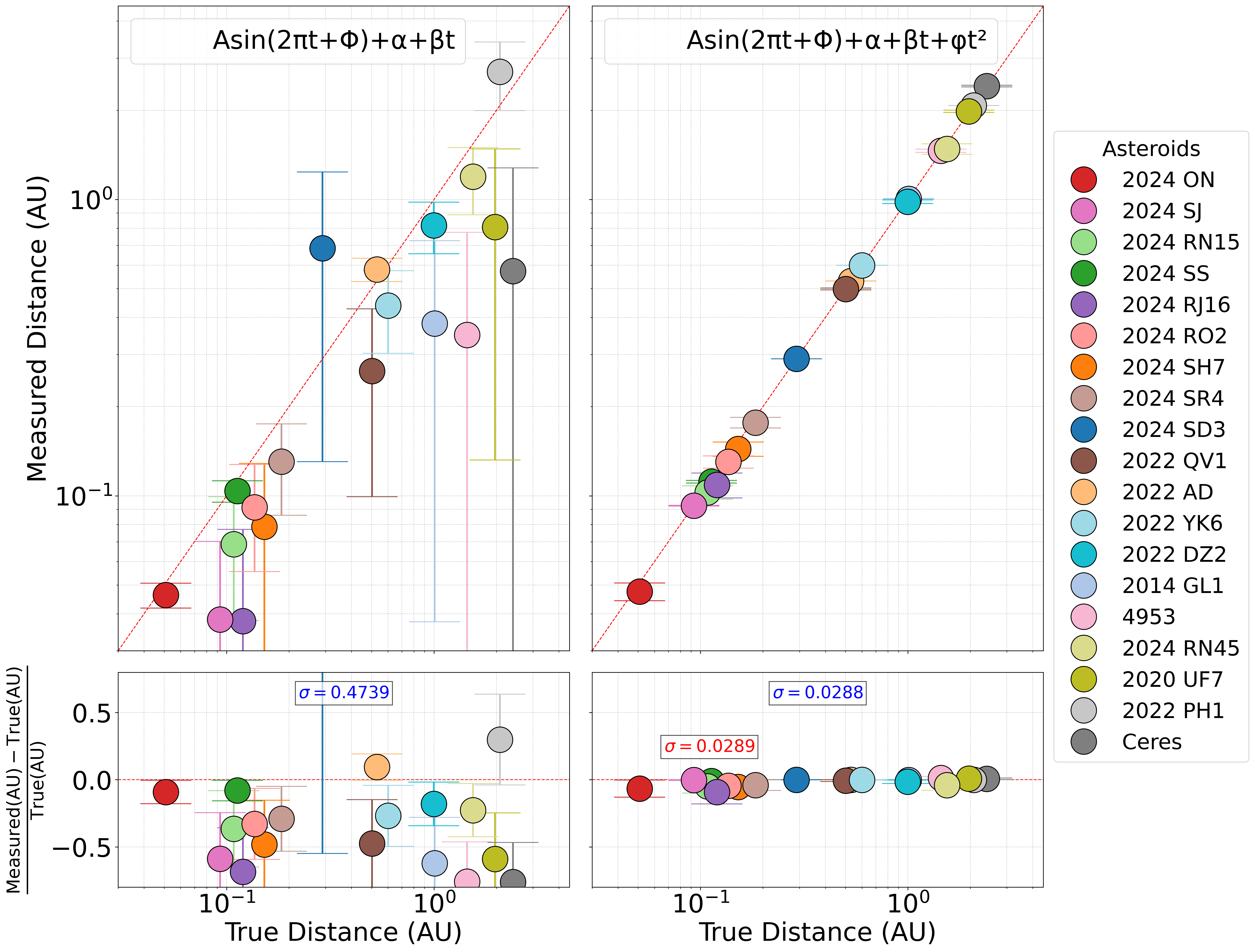}
    \caption{Similar to Fig.~\ref{fig:One_day}, this scenario features 15 observations taken over 5 nights, with three observations per night from September 5-9, 2024, between 20:00:00 and 05:00:00.}
    \noindent\hdashrule[0.5ex]{\linewidth}{0.1mm}{0.1mm}
    \label{fig:Five_days}
\end{figure*}

\subsection{Demonstration of Technique}
\label{sec:section2.1}

In order to better explain the method, we fit our model in steps where we first attempt to isolate the motion of the asteroid itself and then subtract this motion of the asteroid from the overall set of position measurements. As an example, we choose asteroids 2024 ON and 4953/1990 MU (hereafter 4953) since they are the two real-time asteroids we observed and discuss in Section~\ref{sec:section3}. 

As shown in Fig.~\ref{fig:stepstep}; we query positions for asteroid 2024 ON from \textit{Horizons} over 2 days from September 5-6, 2024 at every hour, making up 48 observations. Similarly, as shown in Fig.~\ref{fig:stepstep4953}; we query positions for asteroid 4953 from \textit{Horizons} over two nights from October 30-31, 2024 at every hour, making up 48 observations. In the lower panel of Figs.~\ref{fig:stepstep} and~\ref{fig:stepstep4953}, we subtract out the polynomial portion in the fit of the asteroids motion and show the sinusoidal residual curve, which has a period of 1 day, as expected. From these plots, it is apparent that the asteroid's motion is relatively linear on short time scales and parabolic on longer time scales.  

In practice, rather than following a sequential step as used to demonstrate the process in Figs.~\ref{fig:stepstep} and~\ref{fig:stepstep4953}, we fit the data simultaneously with all parameters as seen in Eqs.~\ref{eq:linear} and~\ref{eq:poly}. While the synthetic data has no uncertainties, we associate uncertainties of $0.1''$ for each observation to be used in Eq.~\ref{eqn:chi^2}. By propagating the amplitude of the sine model through Eqs.~\ref{eq:parallax} and~\ref{eqn:cos}, we measure the distance to 2024 ON at $0.058081 \pm 0.000278$ AU. The synthetic mean distance (true distance) of 2024 ON during the observation window was 0.058442 AU, a deviation of only 0.62\% between measured and synthetic ephemerides data. We note that the distance to 2024 ON itself changed during this observation window at a rate of 0.005 AU per day. This can be observed in the changing of the amplitude at every hour over the course of 2 days. We discuss accounting for this change in the following subsection. Similarly, for 4953, for a synthetic mean distance of 1.14691 AU, we measured the asteroid's distance to be $1.142472 \pm 0.001508$ AU, a 0.39\% deviation. We also note that 4953 changed during this observation window at a rate of 0.00008 AU per day. 

\subsection{Application on Recent NEOs}
\label{sec:Application}

\textit{Horizons} provides accurate synthetic ephemerides data of Solar System Objects including NEOs. We selected nineteen NEOs that covered a range of distances from $\sim0.05 \,\text{AU}$ to $\sim2.4 \,\text{AU}$ from the NASA JPL CNEOS \textit{(NEO Earth Close approaches database)}\footnote{\url{https://cneos.jpl.nasa.gov/ca/}.} and the ESA \textit{(NEOCC Database Statistics)} repository.\footnote{\url{https://neo.ssa.esa.int/search-for-asteroids}.}

Sixteen of these asteroids (2024 SS, 2024 RN15, 2024 RN45, 2024 SJ, 2024 RO2, 2024 SH7, 2024 RJ16, 2024 SR4, 2024 SD3, 2014 GL1, 2020 UF7, 2022 AD, 2022 DZ2, 2022 PH1, 2022 QV1 and 2022 YK6) have absolute \textit{H} $> 24$ \text{mag} (size $\lesssim 90 \, \text{m}$). 4953 and 2024 ON have \textit{H} = 15.0 mag ($\sim 2730$ m) and \textit{H} = 20.5 mag ($\sim210 - 500$ m), respectively. Ceres, a MBA, has \textit{H} = 3.7 mag ($\sim5\times10^{5}$ to $11\times10^{5}$ m). Also, in the synthetic ephemerides data reported on \textit{Horizons} and as noted in the previous section, the error on the positional coordinates is negligible, and so, we fix uncertainties in our calculations at $0.1''$. 

We set the observer's location for all asteroids to `Cerro Tololo Observatory, La Serena, (code: 807)' situated at an altitude of 2200 m and a latitude of 30.1732° S; and input the value of the latitude into Eq.~\ref{eqn:cos}. We study two scenarios, one in which there are six observations with short separations of 45 minutes in a single night \textit{(0.2 days)} from 23:43:00 to 03:28:10 on September 5-6, 2024, and another with 15 observations spread out over 5 days, three observations per night between 20:00:00 and 05:00:00 from September 5-9, 2024. 

In Fig.~\ref{fig:One_day} (column 1), we see the fit from Eq.~\ref{eq:linear} for the group of asteroids yielding precise distances with relatively good agreement with true distances. We define the fractional distance scatter as the root-mean-square (RMS) difference between measured distance ($D_{\mathrm{meas}}$) and true distance ($D_{\mathrm{true}}$) divided by true distance:
\begin{equation}
\label{eq:fractional-distance-scatter}
\text{Fractional Distance Scatter} = 
\mathrm{RMS}\!\left( \frac{D_{\mathrm{meas}} - D_{\mathrm{true}}}{D_{\mathrm{true}}} \right).
\end{equation}. 

We find in the bottom panel (Fig.~\ref{fig:One_day}, column 1), for the distance residuals for all 19 asteroids, the standard deviation of the fractional residuals as defined in Eq.~\ref{eq:fractional-distance-scatter} is $\sigma=0.075$. The $\chi^2$ of the fractional residuals is $19.45$, indicating that the uncertainties for all 19 asteroids in this case are well characterized. The spread of the residuals is only $\sigma=0.0128$ for distances less than 0.3 AU. When Eq.~\ref{eq:poly} is applied (Fig.~\ref{fig:One_day}, column 2), the insertion of the additional non-linear term ($\varphi t^2$) reduces the accuracy of the distance recovery to the asteroids, likely due to the degeneracies between the fit parameters which cannot be broken down by the limited data. A value of $\chi^2\sim50$ also indicates that the uncertainties do not capture the degeneracies in the fit. Therefore, for situations when asteroid data is only within a single night, as discussed later in this paper, we find that Eq.~\ref{eq:linear} is the most optimal function.  

We present another scenario as seen in Fig.~\ref{fig:Five_days} with observations taken over multiple nights. In Fig.~\ref{fig:Five_days} (column 1), Eq.~\ref{eq:linear} produces poor distance agreement; however, Eq.~\ref{eq:poly} which incorporates the polynomial term in Fig.~\ref{fig:Five_days} (column 2), the distance recovery improves and the fractional distance uncertainty is $\sigma = 0.029$ and $\chi^2\sim19.43$. The spread of the residuals for distances less than 0.3 AU is $\sigma = 0.0289$. 

As mentioned for 2024 ON in the previous section, the distance to the asteroid varies throughout the observation window. For the asteroids analyzed here, the distance to Earth changes on the scale of 0.005 AU per 12 hours, accounting for $\sim5\%$ fractional error. To account for this variation, we attempt to modify the sine curve amplitude in Eqs.~\ref{eq:linear} and~\ref{eq:poly} from $\mathcal{A}$ to $\mathcal{A} + \Theta \times t$. We found that the addition of the extra term in the amplitude causes additional degeneracies in the fit and reduced amplitude accuracy; therefore, we do not account for this time-varying parallax amplitude to determine the distances to the asteroids. 

\section{Our Real-time Asteroid Observation and Distance Prediction}
\label{sec:section3}

\begin{figure*}[!htbp]
    \centering
    \includegraphics[width=2.1\columnwidth]{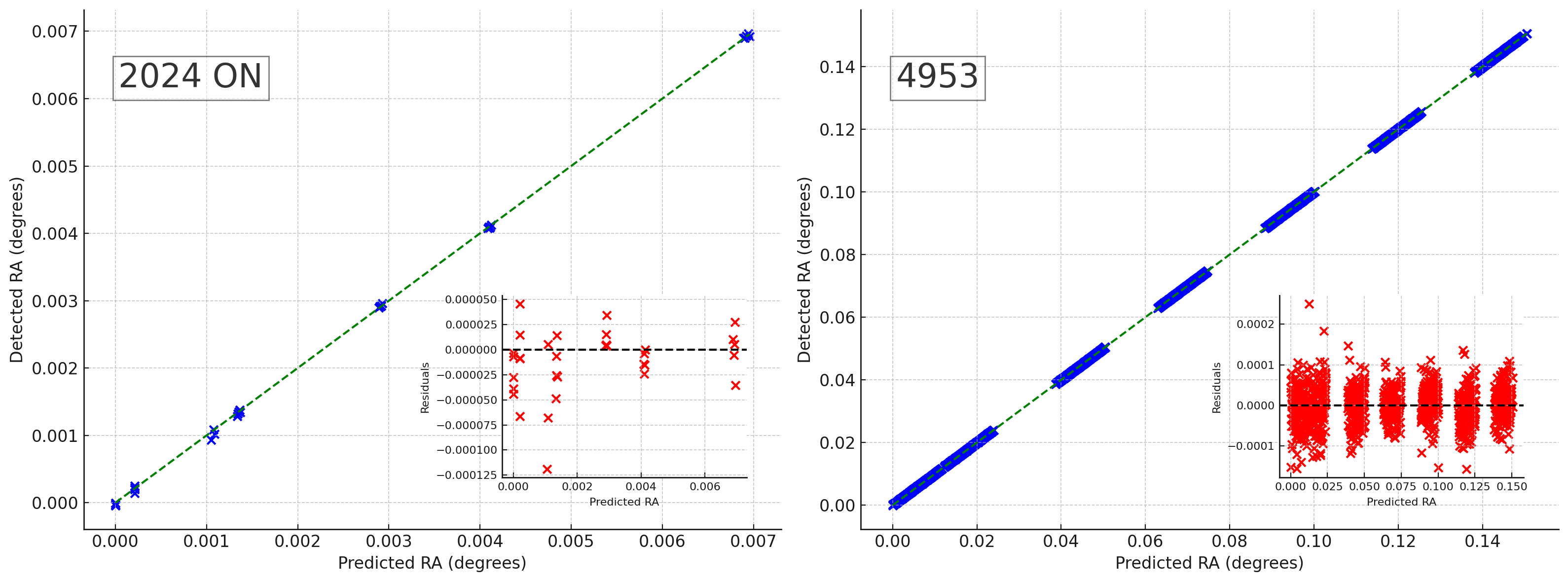}
    \caption{(Main) Measurement of detected RA positions from real-time observations versus predicted RA from \textit{Horizons} for asteroids 2024 ON (left column) and 4953 (right column). We set the initial RA position of the first asteroid detection at 0.0. A line of $y=x$ is overplotted. (Insets) The residuals of the detected RA to the predicted RA. For 2024 ON, seven observations were done in groups of five exposure every $\sim45$ minutes. For 4953, 1049 observations were collected at six different times in groups of multiple exposures per second.}
    \noindent\hdashrule[0.5ex]{\linewidth}{0.1mm}{0.1mm}
    \label{fig:ravra}
\end{figure*}  

To demonstrate the methodology discussed in Section~\ref{sec:section2}, we scheduled a series of real-time observations for asteroid 2024 ON and 4953 with the Panchromatic Robotic Optical Monitoring and Polarimetry Telescopes (PROMPT) on Cerro Tololo in Chile. 

2024 ON was observed for one night on September 5-6, 2024, from 23:43:00 to 03:28:10 at seven different times, with a \textit{V}-band magnitude = 17.6 mag. Each of the seven observations consisted of five exposures, which were co-added, and the mean values of these co-adds is provided in the Appendix. In Fig.~\ref{fig:ravra} (left column), these seven RA observations and their five exposures are plotted with predicted RA from \textit{Horizons}. 

4953 was observed for one night on October 31, from 00:45:14 to 07:24:23. Observations were conducted at six different times, and multiple exposures per second were taken consecutively within each interval, as seen in Fig.~\ref{fig:ravra} (right column). In total, the exposures accounted for 1049 observations. The large number of exposures resulted from a conservative underestimation of the object's brightness. However, 4953 had a \textit{V}-band magnitude = $\sim17.8$ mag, leading to multiple exposures with high signal-to-noise ratios. Unlike 2024 ON, the exposures for 4953 were not co-added due to the large number of evenly spaced exposures. Instead, the asteroid positions were detected in individual exposures and grouped using spline interpolation into eight distinct space/time bins over the observation window, as seen in the appendix. 

We quantify the uncertainty in the observations for both the asteroids using two methods. The first is from the standard deviation (STD) in RA positions from consecutive exposures taken within a short period. For 2024 ON, grouping 5 exposures, we measure STD of $\sim0.08''$. For 4953, grouping 20 exposures taken every minute, we measure STD of $\sim0.15''$. The second method is to measure the STD between detected and synthetic positions. For 2024 ON, the STD is $0.12''$, and for 4953, STD is $0.16''$. We hypothesize that these values are larger than those from individual visits, mainly because time between the visits elapsed, the telescope resettled, and the airmass changed. We use the second set of STD values as it's more conservative and more realistic, though in the future the astrometric uncertainties across the whole night would need to be derived without external sources.

\begin{figure}
    \centering
    \includegraphics[width=\columnwidth, height=10.5cm]{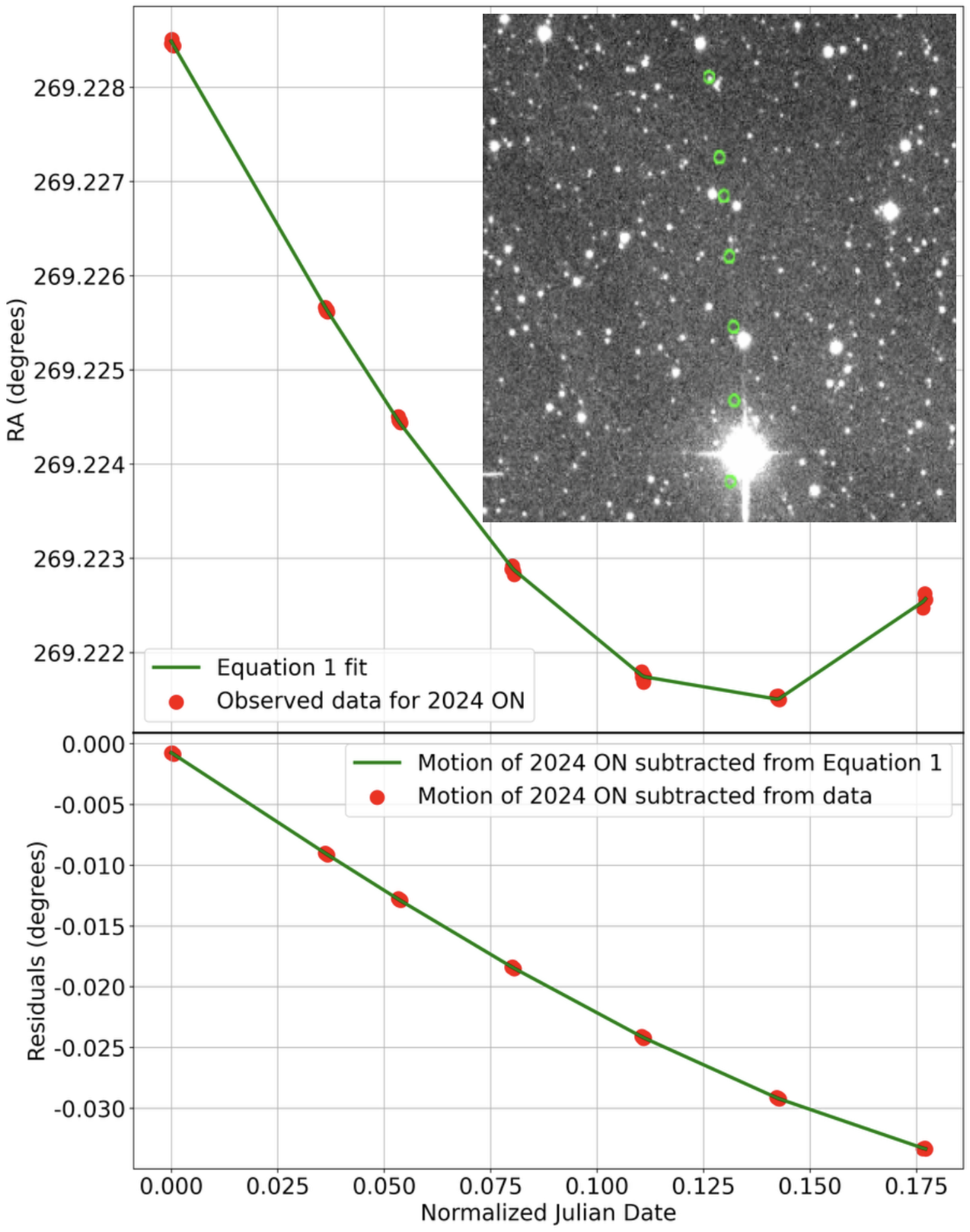}
    \caption{The observations of 2024 ON on Sep 5-6, 2024, between 23:43:00 to 03:28:10. (Top) The fitted curve to the observations is obtained from Eq.~\ref{eq:linear}. (Bottom) Subtraction of the asteroid's linear motion from the overall fit. (Inset) The trail of 2024 ON across the sky for all 7-observations is denoted in green circles.}
    \label{fig:coma}
\end{figure}

\begin{figure}
    \centering
    \includegraphics[width=\columnwidth]{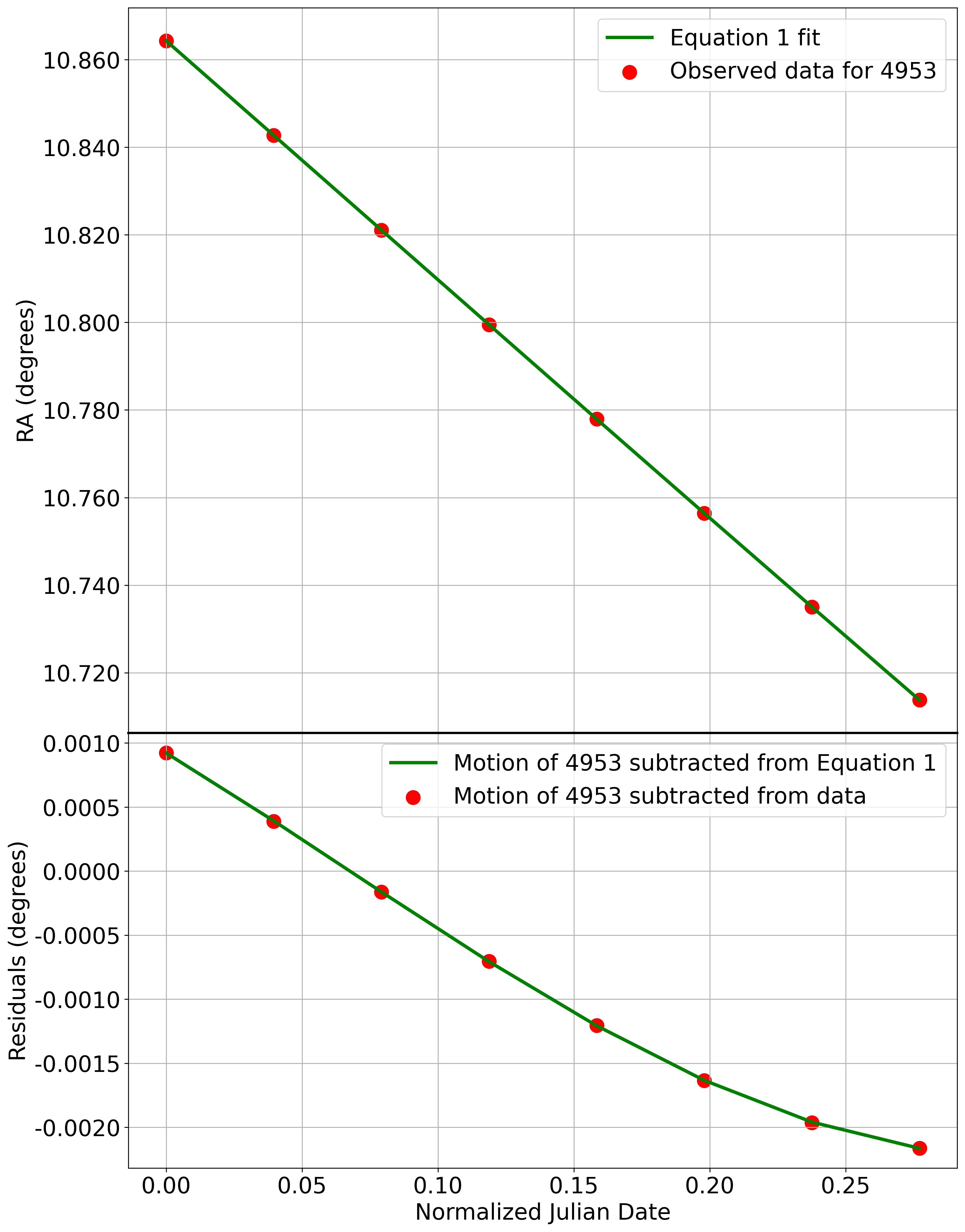}
    \caption{The observations of 4953 on October 31, 2024 from 00:45:14 to 07:24:23. In total, there were 1049 observations which were grouped using spline interpolation into eight distinct bins. (Top) The fitted curve to the observations is from Eq.~\ref{eq:linear}. (Bottom) Subtraction of the asteroid's linear motion from the overall fit.}
    \label{fig:4953_fit}
    \noindent\hdashrule[0.5ex]{\linewidth}{0.1mm}{0.1mm}
\end{figure}

As seen in the insets of Fig.~\ref{fig:ravra}, for both asteroids; 2024 ON has small systematic differences in its exposure groups, different than the per exposure uncertainty, on the order of $\sim0.00001$ degrees, or $\sim0.03''$ between observed and predicted RA. The systematic differences for 4953 is on the order of $\sim0.00003$ degrees or $\sim0.1''$. Amplitudes of the parallax signal for both 2024 ON and 4953 is significantly higher than these systematic effects, making the uncertainties almost negligible. However, for future constraints, we may be able to improve this astrometry, which may be partly due to differential chromatic refraction \citep{Zhai2024}.

In Fig.~\ref{fig:coma}, we show the detection, tracking and centroid determination of 2024 ON, surrounded by many near bright stars and celestial objects. Given our findings in Section~\ref{sec:Application}, Eq.~\ref{eq:linear} is appropriate when observations are within one night. We show the motion of 2024 ON in Fig.~\ref{fig:coma} as well as the best-fit path of the asteroid, and present the results from the fit and the distances measurement in Table~\ref{tab:single_column_table} (column 2). The amplitude (radians) from the sine curve is $0.000643 \pm 0.000013$, and using Eqs.~\ref{eq:parallax} and~\ref{eqn:cos}, we find a parallactic distance of $0.057526 \pm 0.001158$ AU, a $0.78\%$ deviation from 2024 ON's predicted (true) distance (0.057979 AU).

\begin{table*}
\centering
\renewcommand{\arraystretch}{1.3} 
\setlength{\tabcolsep}{13pt} 
\fontsize{11}{11}\selectfont
\begin{tabular}{|l|c|c|c|}
\hline

\multicolumn{3}{|c|}{\textbf{Fit Parameters from Eq.~\ref{eq:linear}}} \\ \hline
\textbf{Parameter} & \multicolumn{1}{|c|}{\textbf{2024 ON}} & \multicolumn{1}{|c|}{\textbf{4953}} \\ \hline

\(\mathcal{A}\) (deg) & 0.036815 $\pm$ 0.000740 & 0.002236 $\pm$ 0.000048 \\ \hline

$\phi$ & 97.40950 $\pm$ 0.012696 & 6.708001 $\pm$ 0.010866 \\ \hline

$\beta$ (deg/day) & 0.150648 $\pm$ 0.005269 & 0.532068 $\pm$ 0.000305 \\ \hline

$\alpha$ (deg) & 269.2292 $\pm$ 0.000459 & 10.86340 $\pm$ 0.000043 \\ \hline

\multicolumn{3}{|c|}{\textbf{Distance Parameters}} \\ \hline
\textbf{Parameter} & \multicolumn{1}{|c|}{\textbf{2024 ON}} & \multicolumn{1}{|c|}{\textbf{4953}} \\ \hline

Amplitude (rad) & 0.000643 $\pm$ 0.000013 & 0.000039 $\pm$ 0.000000 \\ \hline

Parallax distance (km) Eq.~\ref{eq:parallax} & 9915361.29 $\pm$ 199522.00 &  163245295.35 $\pm$ 3531653.61 \\ \hline

Detected distance (km) Eq.~\ref{eqn:cos} & 8605725.45 $\pm$ 173168.83 & 176116983.66 $\pm$ 3810120.10 \\ \hline

Detected distance (AU) & 0.057526 $\pm$ 0.001158 & 1.177269 $\pm$ 0.025469\\ \hline

True distance (AU) & 0.057979 & 1.146913 \\ \hline

Detected distance vs. True distance & 0.78\% & 2.61\% \\ \hline
\end{tabular}

\caption{Fit and distance results of 2024 ON (column 2) and 4953 (column 3). Top table: Fit parameters from Eq.~\ref{eq:linear}. Below table: Distance parameters with uncertainties from Eqs.~\ref{eq:parallax} and~\ref{eqn:cos}.}
\noindent\hdashrule[0.5ex]{\linewidth}{0.1mm}{0.1mm}
\label{tab:single_column_table}
\end{table*}

We repeat this process for asteroid 4953 as seen in Fig.~\ref{fig:4953_fit}, and report the results of the fit and distance measurements in Table~\ref{tab:single_column_table} (column 3). We measure the parallactic distance of 4953 as 1.177269 $\pm$ 0.025469 AU, a 2.61\% deviation from its true distance (1.146913 AU).

\section{Discussion}
\label{sec:section4}

The parallax technique presented in this paper enables rapid and precise distance determination within a single night. We focus on asteroid measurements beyond $0.05-0.1$ AU, as the distance to nearby asteroids can be measured using radar astronomy \citep{Venditti}.  In this section, we characterize additional benefits and limitations of this technique. First, we characterize how well all-sky surveys can measure distances to asteroids post-discovery within a single night. We then discuss limitations of this approach. Finally, we propose modifications to this approach such as following-up on an asteroids motion using multiple 
telescopes at different times, or by using two telescopes at once.

\subsection{Distance Constraints from All-sky Surveys}


There are already a number of all-sky surveys, like CSS, Pan-STARRS, and ATLAS that have found thousands of asteroids.  Upcoming facilities such as the VRO and Argus Array will significantly increase the number of discoveries. Here, we use VRO and Argus as examples to quantify how well distances can be determined within the survey mode for a single night.  

 The VRO in Cerro Pachon in Chile has a primary mirror of 8.4 m diameter, capable of observing 1000 deg$^2$ four times per night and astrometric precision of $0.1''$ to a depth of $r\sim23$ mag \citep{VRO}. The Argus Array is a 900-telescope survey instrument with the equivalent light collecting area of a 5 m telescope, capable of observing 10,000 deg$^2$ eight times per night, with astrometric precision of $\sim0.3''$ to depth of $r\sim21$ mag \citep{Argus}. 
 
We create a toy-model simulation to understand how well the asteroid distances can be constrained with these telescopes over a short observation arc. The characteristics of the asteroids in our simulations are as follows: distances between 0.1 and 0.5 AU (\( \mathcal{A} \) set between 0.00008 and 0.0004, \( \phi \) randomly set between $-1$ and 1), and \( \beta \) set between 0.001 and 0.003 degree/day. The characteristics of the observatory survey is as follows:
astrometric uncertainty from $0.036''$ to $0.18''$; a range of observation windows covering 0.18 days to 0.6 days ($\sim4.3$ hour to $\sim21.9$ hours). We make two cases for the number of exposures over this time range to be 4 and 7 so that surveys such as the LSST have a smaller number of observations and surveys like Argus have more. We note that the distance is significantly more imprecise for less than 4 observations because the distance equation is then unconstrained.  We assume perfect depth in these observations and do not relate distances to astrometric uncertainty. 

We generate ephemerides data using the model given in Eq.~\ref{eq:linear} and add astrometric noise. We re-apply the model to fit and calculate the fractional distance uncertainty for the entire set of asteroids as defined in Eq.~\ref{eq:fractional-distance-scatter}. When doing so, we found that there was a factor of $\sim3\times$ difference in fractional distance uncertainty when there were at least a pair of observations on each side of zenith (when the asteroid is directly overhead), versus when there was not. We give the results assuming zenith is straddled on each side, though note the results can be multiplied by $\sim3$ to see the precision if this condition is not satisfied.

We show the results in Table~\ref{table:scatter-grid}. We find the minimum length of observation window is around 0.15 days or 3.6 hours to constrain the distance well, so begin our table at that value. The fractional distance uncertainty improves with astrometric precision as well as observation-window size. VRO should be able to achieve the lowest levels of astrometric noise as given in the table, and if so, would be able to constrain distances to the sub-percent level. The Argus Array should be able to only reach the higher noise values as given in the table, so we can expect distance constraints at the $3-5\%$ level for observation windows within a single night.  In a future analysis, we will apply the actual survey strategies for these different telescopes to more accurately forecast the distance constraints. Lastly, we also check to see if there are mean biases in the recovered distances, and find that they are constrained to be $10\%$ smaller than the magnitudes of the fractional uncertainty as given in Table~\ref{table:scatter-grid}.

\begin{table}[h!]
\centering
\begin{tabular}{ccccc}
\toprule
\textbf{Noise (deg)} & \multicolumn{4}{c}{\textbf{Observation Window (day)}} \\
\cmidrule(lr){2-5}
 & 0.15 & 0.23 & 0.3 & 0.6  \\
\midrule
4 observations: &&&& \\
$4 \times 10^{-7}$ & 0.0087 & 0.0028 & 0.0019 & 0.0005  \\
$6 \times 10^{-7}$ & 0.0130 & 0.0044 & 0.0029 & 0.0007 \\
$1.25 \times 10^{-6}$ & 0.0271 & 0.0094 & 0.0060 & 0.0015 \\
$2.5 \times 10^{-6}$ & 0.0641 & 0.0191 & 0.0119 & 0.0031  \\
$5 \times 10^{-6}$ & 0.1772 & 0.0371 & 0.0236 & 0.0063  \\
$8 \times 10^{-6}$ & 0.2857 & 0.0633 & 0.0392 & 0.0102  \\
$1 \times 10^{-5}$ & 0.4325 & 0.0801 & 0.0477 & 0.0126  \\

7 observations: &&&& \\
$4 \times 10^{-7}$ &0.0109&0.0036&0.0023&0.0007 \\
$6 \times 10^{-7}$&0.0166&0.0056&0.0037&0.0010 \\
$1.25 \times 10^{-6}$ &0.0387&0.0123&0.0075&0.0020 \\
$2.5 \times 10^{-6}$&0.0949&0.0234&0.0144&0.0040 \\
$5 \times 10^{-6}$&0.2772&0.0514&0.0305&0.0082 \\
$8 \times 10^{-6}$&0.5426&0.0984&0.0499&0.0133 \\
$1 \times 10^{-5}$&0.7423&0.1148&0.0635&0.0164 \\

\bottomrule
\end{tabular}
\caption{Fractional Distance Uncertainty as defined in Eq.~\ref{eq:fractional-distance-scatter} for 4 and 7 evenly spaced observations. The astrometric noise is given in degrees and the separation is given in days. Here we assume there are observations on both sides of zenith.}
\noindent\hdashrule[0.5ex]{\linewidth}{0.1mm}{0.1mm}
\label{table:scatter-grid}
\end{table}

\subsection{Additional Complexities}

While we are able to reach $\sim1\%$ uncertainties as seen in Section~\ref{sec:section3}, we can improve for better astrometric precision, as shown in Table~\ref{table:scatter-grid}. We can also improve for better measurements by considering the following: 

\textbf{Airmass content}: One way to improve astrometric precision and therefore distance accuracy, as described by Z22, is to  schedule observations of the asteroid at low airmass.  When the airmass is high, light from a NEO is diminished/distorted due to atmospheric distortion, scattering, and interferences. At the same time, as shown above with the simulations for future observatories, the distance is best constrained when there is a range of hour angles, preferably straddling zenith. Therefore, one must balance considerations of high/low airmass with needing a range of hour angles.

\textbf{Trailing}: 
Z22 mention that trailing (or blurring) caused by closer NEOs, which appear as long streaks in images, increases astrometric uncertainties and complicates NEO detection and tracking. This astrometric uncertainty degrades signal-to-noise ratio and makes identification and detection of light from far-away objects challenging. Multiple papers (e.g., \citealp{Shao2014,Heinze2015,Zhai2018}, utilize \textit{Synthetic Tracking} and stack multiple short-exposure images in short time frames in rapid succession to establish the predicted trajectory for the fast moving objects. This technique enables the removal of trailing loss and enhances the detection of faint fast-moving NEOs. 

\textbf{Variable Distance During Observation Arc}: As discussed in Section~\ref{sec:Application}, the distance to the asteroid, due to the motion of the asteroid itself may change on the 1-2\% level within a single night. To account for this effect, we attempted to add another parameter though found that our current data was not constraining enough to benefit from this additional parameter.

\subsection{Related Approaches to Measure Distances With Parallax}

While the analysis in this paper focuses on measurements of an asteroid with a single observatory, Z22 also discuss measuring parallax from multiple geographic  baselines for both asynchronous (NEO observed at different times) and synchronous (NEO observed simultaneously) observations. We will pursue this approach in a follow-up work in which we will follow a similar analysis in terms of training on synthetic data, and acquisition of a small sample of asteroid measurements using two different telescopes.

The parallax method discussed in our paper has also been demonstrated using \textit{HST} with asteroid streaks found in archival images \citep{Kruk22}.  The advantages of this approach are: (1) the Hubble Space telescope has a fast orbit of only $\sim84$ minutes at an altitude of $\sim340$ miles in Low-Earth Orbit so there will be a detectable parallax signal within the asteroid streak from a single exposure and (2) \textit{HST} has exquisite astrometric precision reaching the 1 milli-arcsecond scale \citep{Bellini09}. The challenge of this approach is that it is mostly limited to asteroid searches with archival images, and the small field-of-view and limited target-of-opportunity ability of \textit{HST} does not allow follow-up or discovery of large numbers of asteroids. While the parallax signal can be measured with \textit{HST} in Low-Earth Orbit, this signal would not be visible for telescopes like Euclid or the Nancy Grace Roman Space Telescope at the L2 orbit.

A simpler follow-up mode would be to use large ground-based telescopes such as the Large Binoculars Telescope, Gemini Observatory, or Keck Observatory to observe NEOs with \textit{V}-band magnitude = 24--27 mag. This leverages the strengths of large ground-based telescopes in terms of astrometry and depth, and simply requires multiple observations within a single night. One could envision a scenario where LSST discovers asteroids, and these large telescopes with smaller field-of-views gather the follow-up observations for these asteroids.

\section{Conclusions}
\label{sec:conclusion}
In this paper, we demonstrate the feasibility of measuring distances to asteroids in our solar system. We showed how the topocentric parallax technique can retrieve distances to the level of $\sim1.3$\% for distances up to $\sim2.4$ AU. We acquired our own data of two NEOs and retrieved accurate distances to $<1\%$ and $\sim2.5$\%, respectively. We also showed that future telescopes like VRO and the Argus Array can do even better within the normal course of survey operations and potentially within the night of discovery. We showed that the accuracy depends on the range of hour angles in the observations and particularly if observations are taken on both sides of zenith, and that there are a number of ways to improve the astrometric measurements to improve the distance precision.  This approach can be combined with other rapid distance techniques for optimal distance determination.

The code used in the study of this paper is available at: \url{https://github.com/mf342/Maryann-et-al.git}.

\begin{acknowledgements}
\section*{Acknowledgements}
 D.S.~is supported by Department of Energy grant DE-SC0010007, the David and Lucile Packard Foundation, the Templeton Foundation, and Sloan Foundation. 

The observations of asteroids 2024 ON and 4953 were supported by T.L.~NASA grant 80NSSC24K0444.

Maryann would also like to thank her advisor Daniel Scolnic, research group, and family for their unwavering support in the Ph.D.~research journey of her first paper.
\end{acknowledgements}

\newpage

\begin{appendix}

\noindent In Table~\ref{tab:mean_data_predicted}, we report the predicted and the measured positions of 2024 ON over one night on September 5 - 6, 2024 from 23:43:00 to 03:28:10. The predicted observations are taken from \textit{Horizons} and are seen in columns 2 and 3, and the real-time observations as discussed in Section~\ref{sec:section3} and as seen in Figures~\ref{fig:ravra} (left column) and~\ref{fig:ravra} comprise seven observations with five exposures each. These real-time exposures are co-added and their means are provided in columns 4 and 5. 

\begin{table*}[!htbp]
\centering
\renewcommand{\arraystretch}{1.2} 
\setlength{\tabcolsep}{6pt} 
\fontsize{10}{10}\selectfont
\begin{tabular}{|c|c|c|c|c|}
\hline
\multicolumn{5}{|c|}{\textbf{\large 2024 ON: Predicted versus Detected Observations}} \\ \hline
\textbf{MJD} & \textbf{RA Predicted (deg)} & \textbf{DEC Predicted (deg)} & \textbf{RA Detected (deg)} & \textbf{DEC Detected (deg)} \\ \hline
60558.988310 & 269.228500 & 5.079583  & 269.228441 & 5.079513  \\ \hline
60559.024456 & 269.225665 & 5.060008  & 269.225640 & 5.059953  \\ \hline
60559.041609 & 269.224469 & 5.050597  & 269.224479 & 5.050498  \\ \hline
60559.068252 & 269.222912 & 5.035871  & 269.222860 & 5.035864  \\ \hline
60559.098796 & 269.221750 & 5.018794  & 269.221771 & 5.018724  \\ \hline
60559.130498 & 269.221542 & 5.000903  & 269.221510 & 5.000857  \\ \hline
60559.164850 & 269.222592 & 4.981305  & 269.222449 & 4.981453  \\ \hline
\end{tabular}
\caption{The predicted (true) observations from \textit{Horizons} and the detected observations from the PROMPT telescopes at \textit{V}-band magnitude = 17.6 mag. In total, the PROMPT telescope captured seven observations with five exposures each, and the positions using the co-added exposures are provided along with the observation date for asteroid 2024 ON.}
\label{tab:mean_data_predicted}
\noindent\hdashrule[0.5ex]{\linewidth}{0.1mm}{0.1mm} 
\end{table*}

\noindent In Table~\ref{tab:mean_data_predicted_4953}, we report the predicted and the observed positions of 4953 over one night on October 31, 2024 from 00:45:14 to 07:24:23. The real-time observations as discussed in Section~\ref{sec:section3} and as seen in Figures~\ref{fig:ravra} (right column) and~\ref{fig:4953_fit} were taken at six different times, with multiple exposures per second during each interval, totalling 1049 observations. These real-time exposures were spline interpolated for better data fitting into eight distinct bins as seen in columns 4 and 5.

\begin{table*}[!htbp]
\centering
\renewcommand{\arraystretch}{1.2} 
\setlength{\tabcolsep}{6pt} 
\fontsize{10}{10}\selectfont
\begin{tabular}{|c|c|c|c|c|}
\hline
\multicolumn{5}{|c|}{\textbf{\large 4953: Predicted versus Detected Observations}} \\ \hline

\textbf{JD (days)} & \textbf{RA Predicted (deg)} & \textbf{DEC Predicted (deg)} & \textbf{RA Detected (deg)} & \textbf{DEC Detected (deg)} \\ \hline
0.000000  & 10.864287  & -36.743153  & 10.864322 & -36.743048 \\ \hline
0.039598  & 10.842677  & -36.740417  & 10.842719 & -36.741213 \\ \hline
0.079196  & 10.821261  & -36.737686  & 10.821097 & -36.735745 \\ \hline
0.118795  & 10.799497  & -36.734738  & 10.799487 & -36.736984 \\ \hline
0.158393  & 10.777880  & -36.731621  & 10.777918 & -36.731593 \\ \hline
0.197991  & 10.756413  & -36.728536  & 10.756420 & -36.728510 \\ \hline
0.237589  & 10.734989  & -36.725368  & 10.735022 & -36.725451 \\ \hline
0.277187  & 10.713731  & -36.722145  & 10.713754 & -36.722171 \\ \hline

\end{tabular}
\caption{Similar to Table~\ref{tab:mean_data_predicted}. 4953 was also observed using the PROMPT telescopes at \textit{V}-band magnitude = $\sim17.8$ mag, totalling 1049 observations. For better data fitting, the observations were grouped using spline interpolation into eight distinct bins.}
\label{tab:mean_data_predicted_4953}
\end{table*}

\end{appendix}

\vspace{5 cm}

\bibliographystyle{mn2e}
\bibliography{main}{}

\begin{thebibliography}{20}
\providecommand{\natexlab}[1]{#1}
\providecommand{\url}[1]{\texttt{#1}}
\providecommand{\urlprefix}{URL }
\providecommand{\eprint}[1][]{\url{#1}}

\bibitem[{{Alvarez} \& {Buchheim}(2012)}]{Alvarez2012}
{Alvarez}, E.~M., {Buchheim}, R.~K., 2012, Society for Astronomical Sciences Annual Symposium, 31, 45

\bibitem[{{Bellini} \& {Bedin}(2009)}]{Bellini09}
{Bellini}, A., {Bedin}, L.~R., 2009, \pasp, 121, 886, 1419, \eprint arXiv:{0910.3250}

\bibitem[{{Branham}(2005)}]{Branham2005}
{Branham}, R.~L., 2005, Celestial Mechanics and Dynamical Astronomy, 93, 1-4, 53

\bibitem[{{Christensen} et~al.(2018){Christensen}, {Africano} et~al.}]{CSS}
{Christensen}, E., {Africano}, B., {Farneth}, G., et~al., 2018, in AAS/Division for Planetary Sciences Meeting Abstracts \#50, vol.~50 of \emph{AAS/Division for Planetary Sciences Meeting Abstracts}, 310.10

\bibitem[{{Folkner} et~al.(2014){Folkner}, {Williams}, {Boggs}, {Park} \& {Kuchynka}}]{Ryan2014}
{Folkner}, W.~M., {Williams}, J.~G., {Boggs}, D.~H., {Park}, R.~S., {Kuchynka}, P., 2014, Interplanetary Network Progress Report, 42-196, 1

\bibitem[{{Guo} et~al.(2023){Guo}, {Peng}, {Lin} \& {Cao}}]{Guo2023}
{Guo}, B.~F., {Peng}, Q.~Y., {Lin}, F.~R., {Cao}, J.~L., 2023, \aj, 165, 3, 128

\bibitem[{{Heinze} \& {Metchev}(2015)}]{Heinze2015}
{Heinze}, A.~N., {Metchev}, S., 2015, \aj, 150, 4, 124, \eprint arXiv:{1508.06331}

\bibitem[{{Jedicke} et~al.(2015){Jedicke}, {Granvik}, {Micheli}, {Ryan}, {Spahr} \& {Yeomans}}]{NEO's}
{Jedicke}, R., {Granvik}, M., {Micheli}, M., {Ryan}, E., {Spahr}, T., {Yeomans}, D.~K., 2015, in Asteroids IV, edited by {Michel}, P., {DeMeo}, F.~E., {Bottke}, W.~F., 795--813

\bibitem[{{Kruk} et~al.(2022){Kruk}, {Garc{\'\i}a Mart{\'\i}n} et~al.}]{Kruk22}
{Kruk}, S., {Garc{\'\i}a Mart{\'\i}n}, P., {Popescu}, M., et~al., 2022, \aap, 661, A85, \eprint arXiv:{2202.00246}

\bibitem[{{Law} et~al.(2022){Law}, {Corbett} et~al.}]{Argus}
{Law}, N.~M., {Corbett}, H., {Galliher}, N.~W., et~al., 2022, \pasp, 134, 1033, 035003, \eprint arXiv:{2107.00664}

\bibitem[{{Lee} et~al.(2023){Lee}, {Acevedo} et~al.}]{Lee2023}
{Lee}, J., {Acevedo}, M., {Sako}, M., et~al., 2023, \aj, 165, 6, 222, \eprint arXiv:{2304.01858}

\bibitem[{{Shao} et~al.(2014){Shao}, {Nemati} et~al.}]{Shao2014}
{Shao}, M., {Nemati}, B., {Zhai}, C., et~al., 2014, \apj, 782, 1, 1, \eprint arXiv:{1309.3248}

\bibitem[{{Teets} \& {Whitehead}(1999)}]{Teets1999}
{Teets}, D., {Whitehead}, K., 1999, Mathematics Magazine, 72, 2, 83

\bibitem[{{Tonry} et~al.(2018){Tonry}, {Denneau} et~al.}]{ATLAS_asteroids}
{Tonry}, J.~L., {Denneau}, L., {Heinze}, A.~N., et~al., 2018, \pasp, 130, 988, 064505, \eprint arXiv:{1802.00879}

\bibitem[{{Venditti} et~al.(2023){Venditti}, {Marshall}, {Devog{\`e}le}, {Zambrano-Marin} \& {McGilvray}}]{Venditti}
{Venditti}, F. C.~F., {Marshall}, S.~E., {Devog{\`e}le}, M., {Zambrano-Marin}, L.~F., {McGilvray}, A., 2023, Acta Astronautica, 210, 610

\bibitem[{{Vera C. Rubin Observatory LSST Solar System Science Collaboration} et~al.(2020){Vera C. Rubin Observatory LSST Solar System Science Collaboration}, {Jones} et~al.}]{VRO}
{Vera C. Rubin Observatory LSST Solar System Science Collaboration}, {Jones}, R.~L., {Bannister}, M.~T., et~al., 2020, arXiv e-prints, arXiv:2009.07653, \eprint arXiv:{2009.07653}

\bibitem[{{Wainscoat} et~al.(2022){Wainscoat}, {Weryk} et~al.}]{Pan-STARRS}
{Wainscoat}, R., {Weryk}, R., {Ramanjooloo}, Y., et~al., 2022, in AAS/Division for Planetary Sciences Meeting Abstracts, vol.~54 of \emph{AAS/Division for Planetary Sciences Meeting Abstracts}, 504.01

\bibitem[{{Zhai} et~al.(2024){Zhai}, {Shao} et~al.}]{Zhai2024}
{Zhai}, C., {Shao}, M., {Saini}, N., et~al., 2024, \pasp, 136, 3, 034401, \eprint arXiv:{2401.03255}

\bibitem[{{Zhai} et~al.(2018){Zhai}, {Shao} et~al.}]{Zhai2018}
{Zhai}, C., {Shao}, M., {Saini}, N.~S., et~al., 2018, \aj, 156, 2, 65, \eprint arXiv:{1805.01107}

\bibitem[{{Zhai} et~al.(2022){Zhai}, {Shao} et~al.}]{Zhai2022}
{Zhai}, C., {Shao}, M., {Saini}, N.~S., et~al., 2022, \pasp, 134, 1031, 015005

\end{thebibliography}

\end{document}